\documentclass[11pt]{cernart}
\usepackage[dvips]{graphicx}
\usepackage{epsfig}
\usepackage{amsmath,amssymb}
\usepackage[left]{lineno}

\begin{document}
\centerline{\LARGE EUROPEAN ORGANIZATION FOR NUCLEAR RESEARCH}
\vspace{15mm}
{\flushright{
CERN-PH-EP-2012-288\\
\today
\\}}
\vspace{15mm}

\begin{center}
{\bf {\Large \boldmath{ Measurement of the branching ratio of the decay $\Xi^{0}\rightarrow \Sigma^{+} \mu^{-} \overline{\nu}_{\mu}$}}}
\end{center}
\begin{Authlist}
\begin{center}
{\bf NA48/1 Collaboration\footnotemark[1]}
\  \\[0.2cm] 
 %
%
J.R.~Batley,
G.E.~Kalmus\footnotemark[2],
C.~Lazzeroni\footnotemark[3],
D.J.~Munday,
M.~Patel\footnotemark[4],
M.W.~Slater,
S.A.~Wotton \\
{\em \small Cavendish Laboratory, University of Cambridge, Cambridge, CB3 0HE,
UK\footnotemark[5]} \\[0.2cm] 
R.~Arcidiacono\footnotemark[6],
G.~Bocquet,
A.~Ceccucci,
D.~Cundy\footnotemark[7],
N.~Doble\footnotemark[8],
V.~Falaleev,
L.~Gatignon,
A.~Gonidec,
P.~Grafstr\"om,
W.~Kubischta,
F.~Marchetto\footnotemark[9],
I.~Mikulec\footnotemark[10],
A.~Norton,
B.~Panzer-Steindel,
P.~Rubin\footnotemark[11],
H.~Wahl\footnotemark[12] \\
{\em \small CERN, CH-1211 Gen\`eve 23, Switzerland} \\[0.2cm] 
E.~Goudzovski\footnotemark[3],
P.~Hristov\footnotemark[4],
V.~Kekelidze,
V.~Kozhuharov, 
L.~Litov,
D.~Madigozhin,
N.~Molokanova,
Yu.~Potrebenikov,
S.~Stoynev,
A.~Zinchenko\\
{\em \small Joint Institute for Nuclear Research, Dubna, Russian    Federation} \\[0.2cm] 
E.~Monnier\footnotemark[13],
E.C.~Swallow\footnotemark[14],
R.~Winston\footnotemark[15]\\
{\em \small The Enrico Fermi Institute, The University of Chicago, Chicago, Illinois, 60126, U.S.A.}\\[0.2cm]
R.~Sacco\footnotemark[16],
A.~Walker \\
{\em \small Department of Physics and Astronomy, University of    Edinburgh, JCMB King's Buildings, Mayfield Road, Edinburgh,    EH9 3JZ, U.K.} \\[0.2cm] 
%
W.~Baldini,
P.~Dalpiaz,
P.L.~Frabetti,
A.~Gianoli,
M.~Martini,
F.~Petrucci,
M.~Scarpa,
M.~Savri\'e \\
{\em \small Dipartimento di Fisica dell'Universit\`a e Sezione dell'INFN di Ferrara, I-44100 Ferrara, Italy} \\[0.2cm] 
%
%
A.~Bizzeti\footnotemark[17],
M.~Calvetti,
G.~Collazuol\footnotemark[18],
E.~Iacopini,
M.~Lenti,
G.~Ruggiero\footnotemark[4],
M.~Veltri\footnotemark[19] \\
{\em \small Dipartimento di Fisica dell'Universit\`a e Sezione dell'INFN di Firenze, I-50125 Firenze, Italy} \\[0.2cm] 
%
%
M.~Behler,
K.~Eppard,
M.~Eppard,
A.~Hirstius,
K.~Kleinknecht,
U.~Koch,
L.~Masetti, 
P.~Marouelli,
U.~Moosbrugger,
C.~Morales Morales\footnotemark[20],
A.~Peters\footnotemark[4],
R.~Wanke,
A.~Winhart \\
{\em \small Institut f\"ur Physik, Universit\"at Mainz, D-55099 Mainz,
Germany\footnotemark[21]} \\[0.2cm] 
A.~Dabrowski,
T.~Fonseca Martin\footnotemark[4],
M.~Velasco \\
{\em \small Department of Physics and Astronomy, Northwestern University, Evanston Illinois 60208-3112, U.S.A.}
 \\[0.2cm] 
G.~Anzivino,
P.~Cenci,
E.~Imbergamo,
G.~Lamanna\footnotemark[4],
P.~Lubrano,
A.~Michetti,
A.~Nappi,
M.~Pepe,
M.C.~Petrucci,
M.~Piccini\footnotemark[22],
M.~Valdata \\
{\em \small Dipartimento di Fisica dell'Universit\`a e Sezione    dell'INFN di Perugia, I-06100 Perugia, Italy} \\[0.2cm] 
%
%
C.~Cerri,
F.~Costantini,
R.~Fantechi,
L.~Fiorini\footnotemark[23],
S.~Giudici,
I.~Mannelli,
G.~Pierazzini,
M.~Sozzi \\
{\em \small Dipartimento di Fisica, Scuola Normale Superiore e Sezione dell'INFN di Pisa, I-56100 Pisa, Italy} \\[0.2cm] 
%
%
C.~Cheshkov\footnotemark[24],
J.B.~Cheze,
M.~De Beer,
P.~Debu,
G.~Gouge,
G.~Marel,
E.~Mazzucato,
B.~Peyaud,
B.~Vallage \\
{\em \small DSM/DAPNIA - CEA Saclay, F-91191 Gif-sur-Yvette, France} \\[0.2cm] 
M.~Holder,
A.~Maier,
M.~Ziolkowski \\
{\em \small Fachbereich Physik, Universit\"at Siegen, D-57068 Siegen,
Germany\footnotemark[25]} \\[0.2cm] 
S.~Bifani\footnotemark[26],
C.~Biino,
N.~Cartiglia,
M.~Clemencic\footnotemark[4], 
S.~Goy Lopez\footnotemark[27],
E.~Menichetti,
N.~Pastrone \\
{\em \small Dipartimento di Fisica Sperimentale dell'Universit\`a e    Sezione dell'INFN di Torino,  I-10125 Torino, Italy} \\[0.2cm] 
W.~Wislicki,
\\
{\em \small Soltan Institute for Nuclear Studies, Laboratory for High    Energy
Physics,  PL-00-681 Warsaw, Poland\footnotemark[28]} \\[0.2cm] 
H.~Dibon,
M.~Jeitler,
M.~Markytan,
G.~Neuhofer,
L.~Widhalm \\
{\em \small \"Osterreichische Akademie der Wissenschaften, Institut  f\"ur
Hochenergiephysik,  A-10560 Wien, Austria\footnotemark[29]} \\[1cm] 
%
\it{Submitted for publication in Physics Letters B.}
\end{center}
\setcounter{footnote}{0}
\footnotetext[1]{Copyright CERN, for the benefit of the NA48/1 Collaboration}
\footnotetext[2]{Present address: Rutherford Appleton Laboratory,
Chilton, Didcot, OX11 0QX, United Kingdom}
\footnotetext[3]{Present address: School of Physics and Astronomy, The University of Birmingham, Birmingham B15 2TT, United Kingdom}
\footnotetext[4]{Present address: CERN, CH-1211 Gen\`eve 23, Switzerland} 
\footnotetext[5]{ Funded by the U.K.    Particle Physics and Astronomy Research Council}
\footnotetext[6]{Present address: Universit\`a degli Studi del Piemonte Orientale, Via Generale Ettore Perrone 18, 28100 Novara, Italy}
\footnotetext[7]{Present address: Instituto di Cosmogeofisica del CNR di Torino, I-10133 Torino, Italy}
\footnotetext[8]{Also at Dipartimento di Fisica dell'Universit\`a e Sezione dell'INFN di Pisa, I-56100 Pisa, Italy}
\footnotetext[9]{On leave from Sezione dell'INFN di Torino,  I-10125 Torino, Italy}
\footnotetext[10]{ On leave from \"Osterreichische Akademie der Wissenschaften, Institut  f\"ur Hochenergiephysik,  A-1050 Wien, Austria}
\footnotetext[11]{On leave from University of Richmond, Richmond, VA, 23173, 
USA; supported in part by the US NSF under award \#0140230. Present address:
Department of Physics and Astronomy George Mason University, Fairfax, VA 22030A,
USA}
\footnotetext[12]{Also at Dipartimento di Fisica dell'Universit\`a e Sezione dell'INFN di Ferrara, I-44100 Ferrara, Italy}
\footnotetext[13]{Present address: Centre de Physique des Particules de Marseille, IN2P3-CNRS, Universit\'e 
de la M\'editerran\'ee, Marseille, France}
\footnotetext[14]{Present address: Department of Physics, Elmhurst College, 
Elmhurst, IL, 60126, USA}
\footnotetext[15]{Also at University of California, Merced, USA}
\footnotetext[16]{Present address: Department of Physics Queen Mary, University
of London, Mile End Road, London E1 4NS, United Kingdom}
\footnotetext[17]{ Dipartimento di Fisica dell'Universit\`a di Modena e Reggio Emilia, via G. Campi 213/A I-41100, Modena, Italy}
\footnotetext[18]{Present address: Dipartimento di Fisica E Astronomia "Galileo Galilei", 
Universit\`a degli Studi di Padova, via 8 febbraio 2, 35122 Padova, Italy} 
\footnotetext[19]{ Istituto di Fisica, Universit\`a di Urbino, I-61029  Urbino, Italy}
\footnotetext[20]{Present address: Institut f\"ur Kernphysik, Universit\"at Mainz, D-55099 Mainz, Germany}
\footnotetext[21]{ Funded by the German Federal Minister for    Research and Technology (BMBF) under contract 7MZ18P(4)-TP2}
\footnotetext[22]{\noindent contact: Mauro.Piccini@pg.infn.it}
\footnotetext[23]{Present address: Universidad de Valencia - Instituto de F\'isica Corpuscular (IFC)
Edificio Institutos de Investigaci\'on, c/ Catedr\'atico Jos\'e Beltr\'an, 2, E-46980 Paterna, Spain}
\footnotetext[24]{Present address: Universit\'e de Lyon, Universit\'e Lyon 1, CNRS/IN2P3, IPN-Lyon, Villeurbanne, France} 
\footnotetext[25]{Present address: School of Physics, Science Centre North, University College Dublin, Belfield, Dublin 4, Ireland} 
\footnotetext[26]{ Funded by the German Federal Minister for Research and Technology (BMBF) under contract 056SI74}
\footnotetext[27]{Present address: CIEMAT - Centro de Investigaciones Energ\'eticas, Medioambientales y 
Tecnol\'ogicas, Avda. Complutense, 22, E-28040 Madrid, Spain} 
\footnotetext[28]{Supported by the Committee for Scientific Research grants
5P03B10120, SPUB-M/CERN/P03/DZ210/2000 and SPB/CERN/P03/DZ146/2002}
\footnotetext[29]{Funded by the Austrian Ministry for Traffic and 
Research under the 
contract GZ 616.360/2-IV GZ 616.363/2-VIII, 
and by the Fonds f\"ur   Wissenschaft und Forschung FWF Nr.~P08929-PHY}

\end{Authlist}


%
%
%
%
%
%
\newpage
\begin{abstract}

From the 2002 data taking with a neutral kaon beam extracted from the CERN-SPS, the NA48/1 experiment observed  
97 $\Xi^{0}\rightarrow \Sigma^{+} \mu^{-} \overline{\nu}_{\mu}$ candidates 
with a background contamination of $30.8 \pm 4.2$ events.
 From this sample, the BR($\Xi^{0}\rightarrow \Sigma^{+} \mu^{-} \overline{\nu}_{\mu}$) is measured to be $(2.17 \pm 0.32_{\mathrm{stat}}\pm 0.17_{\mathrm{syst}})\times10^{-6}$.
\end{abstract}

\newpage



\vspace{0.2cm}

\section{Introduction}
The study of hadron beta decays gives important information on the interplay 
between the weak interaction and hadron structure determined by the strong 
interaction.
In this framework, measurements on $\Xi^{0}$ semileptonic decays and on the 
related parameters are fundamental to further increase our knowledge on the 
constituents of the baryon octet.
In particular, a clear evidence for the decay $\Xi^{0}\rightarrow \Sigma^{+} \mu^{-} \overline{\nu}_{\mu}$ and a measurement of its branching ratio
will add one more constraint to the theoretical 
frameworks~\cite{donoghue,mendieta,schlumpf,krause,anderson} built to explain
the behaviour of the baryon semileptonic decays.

In the present article the branching ratio of the semileptonic decay  
$\Xi^{0}\rightarrow \Sigma^{+} \mu^{-} \overline{\nu}_{\mu}$ is measured by normalizing to the analogue
decay with an electron in the final state, already studied by the NA48/1 collaboration~\cite{electron}. The similar topologies of the final states of the two semileptonic decays
allowed the same trigger conditions to be used for both data samples. The selection criteria are also
similar between the two channels and only differ for the identification of the charged lepton and 
for cuts related to background rejection.
This decay had already been observed by the KTeV collaboration~\cite{ktevmu}, with a sample of 9 events and a branching ratio measurement of 
$(4.7 ^{+2.2}_{-1.6})\times 10^{-6}$.

\section{Beam}
The experiment was performed in 2002 at the CERN SPS accelerator and used a
400~GeV proton beam impinging on a Be target 
to produce a neutral beam. The spill length was 4.8~s out of a 16.2~s 
cycle time. The proton intensity was fairly constant during the spill with a 
mean of $5 \times 10^{10}$  particles per pulse.

%

For this measurement, only
the $K_{S}$ target station of the NA48 double 
$K_{S}/K_{L}$ beam line~\cite{detector} 
was used to produce the neutral beam. In this configuration,
the $K_L$ beam was blocked and an additional sweeping magnet 
was installed to deflect charged particles away from 
the defining section of the $K_S$ collimators.
To reduce the number of photons in the neutral beam originating
primarily from $\pi^0$ decays, a 24~mm thick platinum absorber was placed in the 
beam between the target and the collimator.
A pair of coaxial collimators, having a total thickness of 5.1~m, the axis 
of which formed an angle of 4.2~mrad
to the proton beam direction, selected a beam of 
neutral long-lived particles ($K_S$, $K_L$,
$\Lambda^0$, $\Xi^0$, $n$ and $\gamma$). 
The target position and the production angle were chosen in such a way
that the beam axis was hitting the center of the electromagnetic calorimeter.  
  
In order to minimize the 
interaction of the neutral beam with air,
the collimator was immediately followed by a $90$~m long 
evacuated tank terminated by a 0.3$\%$ $X_0$ thick Kevlar 
window. The NA48 detector was located downstream of this region.

On average,  about $1.4 \times 10^{4}$  $\Xi^{0}$ per spill, with an energy 
between 70 and 220~GeV, decayed in
the fiducial decay volume.

\section{Detector}

The detector was designed for the measurement of $Re(\epsilon^{\prime}
/ \epsilon)$, and a detailed description of the experimental layout is available at~\cite{detector}. 
In the following sections a short description of the main detectors
is reported.

\subsection{Tracking}

The detector included a spectrometer 
housed in a helium gas volume
with two drift chambers before and two after
a dipole magnet with an horizontal transverse momentum kick
of 265 MeV/{\it c}. 
Each chamber had four views ({\it x, y, u, v}), each of which had two 
sense wire planes.
The resulting space points were typically reconstructed with a
resolution of $\sim 150$~$\mu$m in each projection.
The spectrometer momentum resolution is parameterized as:

\begin{equation*}
\sigma_p /p = 0.48 \% \oplus 0.015\% \times p 
\end{equation*}

\noindent where $p$ is in GeV/{\it c}. This
gave a resolution of 3 MeV/{\it c}$^2$ 
when reconstructing the kaon mass in $K^{0}\rightarrow \pi^{+}\pi^{-}$\
decays. The track time resolution was $\sim1.4$~ns.

\subsection{Electromagnetic Calorimetry}

The detection and measurement of the electromagnetic showers 
were achieved with a
liquid krypton calorimeter (LKr), 27 radiation lengths deep,  with a 
$\sim$2~cm $\times $ 2~cm cell cross-section.

The energy resolution, expressing $E$ in GeV, 
is parameterized as~\cite{detector}:

\begin{equation*}
\sigma(E) / E = 3.2\% /\sqrt{E} \oplus 9 \% / E \oplus 0.42 \%
\end{equation*}

The transverse position resolution for a single photon
of energy larger than 20 GeV was better than 1.3 mm, and
the corresponding mass resolution for the reconstructed $\pi^0$ mass 
($\gamma \gamma$ decay) was $\sim$1 MeV/{\it c}$^2$.
The time resolution of the calorimeter for a single shower was
better than $\sim300$~ps.

\subsection{Scintillator Detectors and Muon Detector}

A scintillator hodoscope (CHOD) was located between the
spectrometer and the calorimeter. It consisted of two planes, segmented
in horizontal and vertical strips and arranged in four quadrants. 
Further downstream there was an iron-scintillator sandwich 
hadron calorimeter (HAC), followed by muon counters consisting of three planes 
of scintillator, each shielded by an 80~cm thick iron wall.  
The first two planes $M1X$ and $M1Y$ were the main muon counters and 
 had 25~cm wide horizontal and vertical scintillator
strips  respectively,  with a length of 2.7~m.  The third plane, $M2X$,  had
horizontal strips 44.6~cm wide, and was mainly used to measure 
the efficiency of the $M1X$ and $M1Y$ counters.
The central strip in  each plane was divided into two sections separated by a gap 
of 21~cm, in order to accommodate the beam pipe.
The fiducial volume of the experiment was principally determined by the 
LKr calorimeter acceptance, together with 
seven rings of scintillation counters (AKL) 
used to veto activity outside this region.

\section{Trigger}

The trigger system used for the on-line selection of $\Xi^{0}$ semileptonic decays
mainly consisted of two levels of logic. Level 1 (L1) was based on logic 
combinations of fast signals coming from various sub-detectors. It required  
hits in the CHOD and in the first drift chamber compatible with at least one 
and two tracks respectively, no hit in the AKL veto system and a minimum 
energy deposition in the calorimeters. This last requirement was 
15~GeV for the energy reconstructed in the LKr calorimeter or 30~GeV
for the summed energy in the electromagnetic and hadronic calorimeters.
The output rate of the L1 stage was about 50~kHz. The average L1 
efficiency, measured
with $\Xi^0 \rightarrow \Lambda \pi^0$ events of energy greater than 
70~GeV, was found to be $98.65 \pm 0.03\%$.

Level 2 (L2) consisted of a set of 300~MHz processors that reconstructed
 tracks and vertices from hits in the drift chambers 
and computed relevant physical quantities. The L2 trigger required
at least two tracks with a closest distance of approach of less than 8~cm 
in space and a transverse separation greater than 5~cm in the 
first drift chamber. Since the signature of the $\Xi^0$ $\beta$-decay
involves the detection of an energetic proton from the 
subsequent $\Sigma^+ \rightarrow p\pi^0$ decay, the ratio between the higher 
and the lower of the two track momenta was required to be larger 
than 3.5. Rejection of the overwhelming $\Lambda \rightarrow p\pi^-$ and 
$K_S \rightarrow \pi^{+}\pi^{-}$ decays was 
achieved by applying stringent invariant mass cuts
against these decays. The output L2 trigger rate was 
about 2.5~kHz. The efficiency of the L2 trigger stage with respect to 
Level 1, averaged over the 2002 run, was measured to 
be $(83.7 \pm 2.2)\%$ for $\Xi^{0}$ $\beta$-decays, mainly
limited by wire inefficiencies in the drift chambers. 

\section{Offline selection}

The identification of the $\Xi^0 \to \Sigma^+ \mu^- \overline{\nu}_\mu$
channel was performed using the subsequent decay 
$\Sigma^+ \to p \pi^0$ with $\pi^0 \to \gamma \gamma$. 
The final state consists of a proton and a muon, giving two 
tracks in the spectrometer, two photons producing 
clusters in the LKr calorimeter and one unobserved anti-neutrino. 
The decay $\Xi^0 \to \Sigma^+ \ell^- \overline{\nu}_{\ell}$ is the only source 
of $\Sigma^+$ particles in the neutral beam since the two-body decay 
$\Xi^0 \to \Sigma^+ \pi^{-}$ is kinematically forbidden. Thus, the 
signal events were 
identified by requiring an invariant $p\pi^0$ mass consistent with 
the nominal $\Sigma^+$ mass value.

The $\Sigma^+$ decay was reconstructed using a positive charged track 
in the spectro\-meter (associated to the proton) and two clusters in the electromagnetic 
calorimeter (associated to the photons from $\pi^0\rightarrow \gamma \gamma$ decay) 
within a time window of 2~ns.
The longitudinal position of the $\Sigma^+$ decay vertex 
was determined using the $\pi^0$ mass constraint to calculate
the distance of its decay point from the calorimeter:
\begin{eqnarray}
\Delta z_{\pi^0} = \frac{1}{m_{\pi^0}} \sqrt{E_1 E_2 r_{12}^2} 
\end{eqnarray}
where $E_1$ and $E_2$ are the measured energies of the two clusters and
$r_{12}$ is the distance between the two clusters in the transverse plane.
Good candidates were kept if the reconstructed  $p\pi^0$ invariant mass was
within 6~MeV/{\it c}$^2$ of the nominal $\Sigma^+$ mass value.
The mass interval was tightened from 8~MeV/{\it c}$^2$ to 6~MeV/{\it c}$^2$
with respect to the normalization channel (see below) to reduce the higher 
background contamination in the muon channel.

Muon identification was achieved by requiring the presence of in-time 
signals from the first two planes of the muon detector ($\pm$2~ns
with respect to the time measured in the charged hodoscope).
In addition, to reject pions and electrons,
the energy deposited in the electromagnetic calorimeter
in association to the muon track was required to be less than  2.5~GeV.

The lower momentum threshold for the muon track was set to 
7~GeV/{\it c} (it was 4~GeV/{\it c} for the electron channel) to reduce the background contamination and to increase the efficiency
for muon reconstruction (see section \ref{measurement}).

The muon momentum calculated in the $\Sigma^+$
 rest frame was required to be less than 0.125~GeV/{\it c},
exploiting the fact that no contribution is expected
from the signal sample above this limit. This cut was not applied in the normalization channel.
Similarly, since the proton momentum in the signal sample is mostly above 
54~GeV/{\it c}, this criterion was used to enhance the probability that sufficient energy is deposited 
in the electromagnetic and hadron calorimeters to satisfy the trigger 
condition $E_{\mathrm{HAC+LKr}}> 30$~GeV. In the normalization channel the lower cut on the proton momentum was set at 40~GeV/{\it c}. 

The $\Xi^0$ decay vertex position was obtained by computing the 
closest distance of approach between the extrapolated $\Sigma^+$ 
line-of-flight and 
the muon track. This distance was required to be less than 4~cm. 
Furthermore, the deviation of the transverse $\Xi^0$ vertex position from 
the nominal line-of-flight defined by a straight line going from 
the center of the $K_{S}$ target to the 
center of the 
liquid krypton calorimeter was required to be less 
than 3~cm.

The longitudinal position of the $\Xi^0$ vertex was required to be 
at least 6.5~m downstream of the $K_{S}$ target, i.e. 0.5~m after 
the end of the final collimator and at most 40~m from the target. Similarly, the $\Sigma^+$ vertex position was 
required to be at least 6.5~m downstream of the target but at most 50~m 
from the target. The latter value was chosen larger than the upper limit 
for the $\Xi^0$ vertex position to account for the lifetime 
of the $\Sigma^+$ particle. The longitudinal separation between the 
$\Xi^0$ and $\Sigma^+$ decay vertices was required to be between $-8$~m and 
$40$~m. The negative lower limit, tuned with Monte Carlo events, was chosen 
such as to take properly into account resolution effects.

The quantity $\vec{r}_{\mathrm{COG}}$
was defined as $\vec{r}_{\mathrm{COG}} = \sum_i \vec{r}_{i} E_i / \sum_i E_i$ 
where $E_i$ is the energy of the detected particle and $\vec{r}_{i}$   
the corresponding transverse position vector at the 
liquid krypton calorimeter position $z_{LKr}$. For a charged particle, 
the quantity $\vec{r}_{i}$ was obtained from the extrapolation to $z_{LKr}$ of 
the upstream segment of the associated track.
For kinematical reasons, the missing transverse momentum ($p_{t}$) is smaller in the muon case with
 respect to the electron case.
Therefore $\vec{r}_{\mathrm{COG}}$ was required to be less than 8~cm 
instead of 15~cm as for the electron channel.

By requiring the invariant mass $\pi^+ \pi^0 \mu^-$ to be less than
0.490~GeV$/{\it c}^2$, the contamination from
 $K_L \rightarrow \pi^+ \pi^- \pi^0$, when the $\pi^-$ is misidentified
as a muon, was reduced to a negligible level. This cut was not applied in the normalization channel.

Cuts were also applied on the positions of the hit points of the tracks in the chambers and 
on the cluster positions in the electromagnetic calorimeter 
to improve the trigger and the reconstruction efficiencies.
Furthermore the energies of the photons coming from the $\pi^{0}$ decay
were requested to be between 3 and 100~GeV to ensure linearity on the LKr 
measurement.

\begin{figure}[htbp]
  \centering
    \includegraphics[width=12cm]{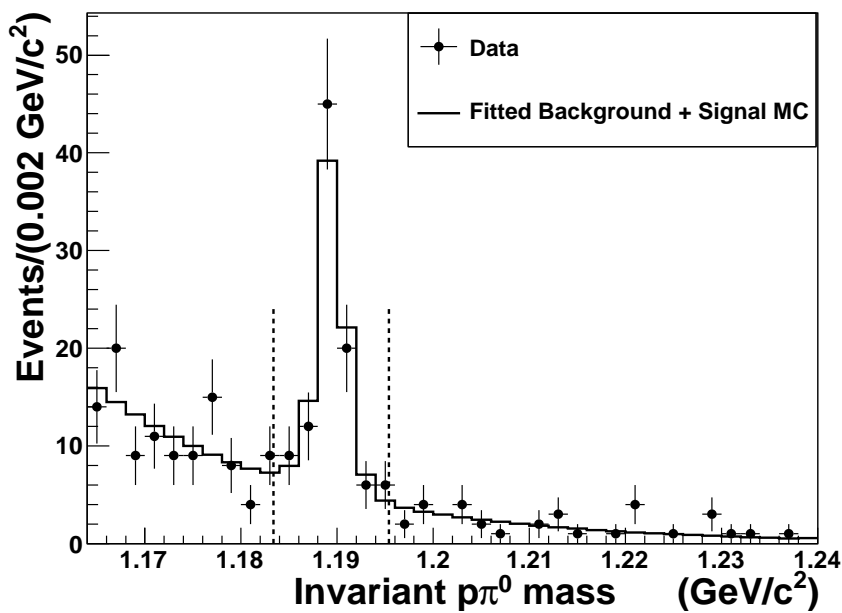}
  \caption{Reconstructed p$\pi^{0}$ invariant mass distribution 
for $\Xi^{0} \rightarrow \Sigma^{+} \mu^{-} \overline{\nu}_{\mu}$ candidates 
after all selection criteria were applied. Points with error bars are data.
The peak at the $\Sigma^{+}$ mass value shows clear evidence for the signal.
The vertical dashed lines delimit the signal region.
The background was evaluated
with a likelihood fit of the data performed in two intervals, 
between 1.164 and 1.180~GeV$/{\it c}^{2}$ and between 1.198 and 1.240~GeV$/{\it c}^{2}$. The solid histogram shows 
the sum of the background contribution (evaluated from the fit) and the Monte Carlo sample of the signal normalized 
to the events found in the data after background subtraction.}
  \label{signal}
\end{figure}

With the above selection criteria, 97  
$\Xi^0 \rightarrow \Sigma^+ \mu^- \overline{\nu}_\mu$ candidates  
were observed in the signal region. The distribution of events in 
the $p\pi^0$ invariant mass variable 
is shown in Figure \ref{signal} after all selection cuts were applied.
Signal events peaking around the $\Sigma^+$ mass are clearly visible
above the background.


A contribution to the background (about $20\%$ of the total) comes from 
overlapping events in the detector (accidentals).
This contribution was estimated directly from data samples, looking to the activity in the detectors 
not in time with the main event time.
There is a small contribution to the background from the decay $\Xi^0\rightarrow \Lambda \pi^0$  
(populating the left side of the $p-\pi^0$ invariant mass distribution) 
with  $\Lambda \rightarrow p \pi^-$ and with $\pi^-$ either misidentified as a muon or 
decaying into $\pi^- \rightarrow \mu^- \nu_{\mu}$.
This contribution was estimated by the Monte Carlo simulation.
However the main contribution to the background is due to scattered events in
the final collimator of the neutral beam, in analogy to what was seen in the 
normalization channel. 

Due to the difficulty to simulate this contribution and due to the low statistics in the control samples coming from data, 
the background distribution in the $p\pi^0$ invariant mass was fitted with an exponential
in the intervals 1.164 - 1.180~GeV$/{\it c}^{2}$ and  1.198 - 1.240~GeV$/{\it c}^{2}$. 
The fit was giving an estimate of $(30.8 \pm 3.8_{\mathrm{stat}} \pm 1.9_{\mathrm{syst}})$  
background events when extrapolated into the signal region.
The systematic uncertainty on the background level was estimated by varying the fitting function and the fit region (also including the signal region, with the signal fitted with a Gaussian distribution).

The data sample for the normalization channel $\Xi^{0} \rightarrow \Sigma^{+} e^{-} \overline{\nu}_{e}$ consists of 6316 events with a background of $(3.4 \pm 0.7)\%$.

A detailed description of the reconstruction and selection for the normalization channel is reported in~\cite{electron}. For that decay, since the electron is completely absorbed in the LKr, the corresponding track was identified by requiring a ratio between the energy deposit in the LKr and the momentum measured by the spectrometer (E/p) greater than 0.85 and lower than 1.15. The other differences in the selection criteria of the signal and normalization channels are described above.

\section{Acceptance}

The acceptance for both signal and normalization 
decay channels was computed using a detailed Monte Carlo program
based on GEANT3~\cite{detector,geant3}. Particle interactions in the detector material 
as well as the response functions of the different detector elements were 
taken into account in the simulation. 
A detailed description of the generator of the electron channel can be found 
in~\cite{electron}.
The generator for the muon channel was modified to include the contribution
from pseudo-scalar currents~\cite{Uli}, parameterized with the
form factor $g_{3}$ which, under Partially Conserved Axial Current (PCAC) 
hypothesis, can be extracted at $q=0$ from the Goldberg-Treiman relation~\cite{Gold,Comm}:
\begin{equation}
g_{3}(0)/f_{1}(0)=2(M_{\Xi^{0}}/M_{K^-})^2 g_{1}(0)/f_{1}(0).
\end{equation} 
Since the $g_3$ term is multiplied by 
$m_{\mathrm{lepton}}/m_{\Xi^{0}}$, its contribution is non-negligible for the muon case~\cite{Uli}.
Using the available experimental results~\cite{electron, ktevff, PDG} for the electron channel, the best estimates 
for the remaining non-vanishing form factors are:
\begin{eqnarray}
   f_{2}(q^{2}=0)/f_{1}(q^{2}=0) &=& 2.0   \pm 1.3 \cr
   g_{1}(q^{2}=0)/f_{1}(q^{2}=0) &=& 1.21  \pm 0.05. 
\label{valoria0}
\end{eqnarray}
The central values were plugged into the Monte Carlo generator and the corresponding errors 
were used  to evaluate the systematic error related to the acceptance 
calculation.   
Radiative corrections were not included in the generator of the muon channel. This leads to 
a systematic uncertainty of 1$\%$, estimated using the Monte Carlo simulation for the electron
channel with the electron mass substituted by the muon one.
The acceptance for the signal $\Xi^{0} \rightarrow \Sigma^{+} \mu^{-} \overline{\nu}_{\mu}$
was calculated to be (3.17 $\pm$ 0.01)$\%$, while 
the acceptance for the normalization $\Xi^{0} \rightarrow \Sigma^{+} e^{-} \overline{\nu}_{e}$ 
was (2.49 $\pm$ 0.01)$\%$. Both quoted uncertainties originate 
from the statistics of the Monte Carlo samples.

\section{$\Xi^0 \rightarrow \Sigma^+ \mu^- \overline{\nu}_{\mu}$ branching ratio}
\label{measurement}
The $\Xi^0 \rightarrow \Sigma^+ \mu^- \overline{\nu}_{\mu}$ 
branching ratio was obtained from the background-subtracted 
numbers of selected events for signal and normalization, the
corresponding acceptance values,     
the normalization branching ratio~\cite{electron} and the efficiency on muon identification. These quantities 
are summarized in Table \ref{brcalc} and yield: 
\begin{eqnarray}
\mathrm{BR}(\Xi^0 \rightarrow \Sigma^+ \mu^- \overline{\nu}_{\mu}) =(2.17 \pm 0.32_{\mathrm{stat}}\pm 0.17_{\mathrm{syst}})\times10^{-6},
\end{eqnarray}
where the statistical uncertainty originates from the event statistics and 
the systematic one is the sum in quadrature of the 
various contributions presented in Table \ref{systematics}.

\begin{table}[htbp]
\begin{center}
\caption{Parameters used for the 
BR($\Xi^0 \rightarrow \Sigma^+ \mu^- \overline{\nu}_{\mu}$) measurement. The numbers 
used for the normalization channel are taken from reference~\cite{electron}. 
\label{brcalc}}
\begin{tabular}{| l | c | c|}
\hline
& $\Xi^0 \rightarrow \Sigma^+ \mu^- \overline{\nu}_{\mu}$ &
$\Xi^0 \rightarrow \Sigma^+ e^- \overline{\nu}_e$ \\
\hline
 Event statistics   & $97$ & $6316$ \\
 Background         & $(30.8\pm4.2)$ events  & $(3.4 \pm 0.7)\%$ \\
 Acceptance         &  $(3.17 \pm 0.01)\%$ & $(2.49\pm0.01)\%$ \\
 Muon inefficiency  &  $(1.5 \pm 0.5)\%$ & \\
 Branching ratio     & & $(2.51 \pm0.03_{\mathrm{stat}}\pm 0.09_{\mathrm{syst}})\times10^{-4}$ \\
\hline
\end{tabular}
\end{center}
\end{table}

The largest contribution to the total systematic uncertainty 
comes from the background subtraction, described above.

From the systematic uncertainty related to the measurement of the 
branching ratio of the normalization channel, the trigger efficiency 
contribution was eliminated, since it is common to both channels. 
A further systematic of $3\%$ was added to take into account the
dependence of the trigger efficiency on the lepton momentum. 

The sensitivity of the branching ratio measurement to the form factors
was studied by varying $g_1(0)/f_1(0)$ and  $f_2(0)/f_1(0)$ within the limits
provided by their uncertainties and doubling or neglecting the $g_3(0)$ value.
\begin{figure}[htbp]
  \centering
    \includegraphics[width=12cm]{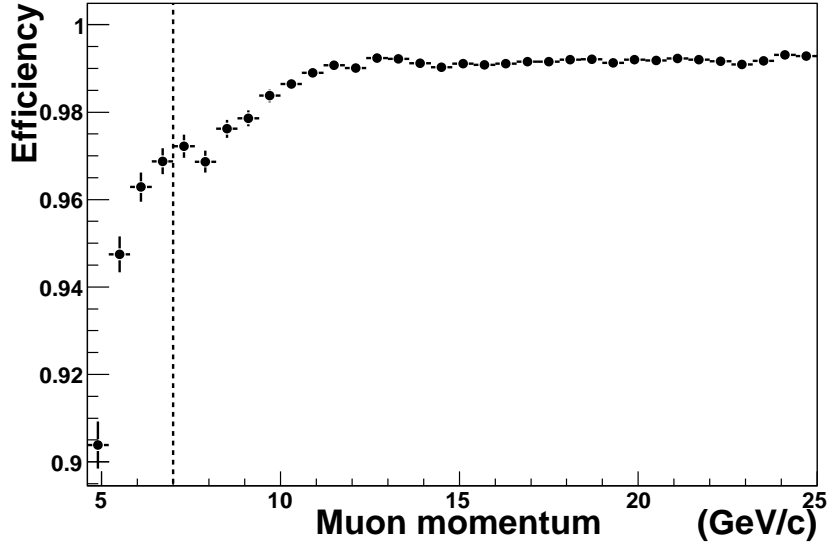}
  \caption{Efficiency for muon identification as a function of muon momentum (right),
the dashed line shows the lower threshold at 7~GeV/{\it c} applied to muon momentum.}
  \label{mueff}
\end{figure}
The muon momentum distribution from Monte Carlo simulation was divided
by the distribution of muon efficiency as a function of muon momentum
as measured from $K^{\pm} \rightarrow \mu^{\pm} \nu_{\mu}$ $(K\mu2)$ decays,
obtained from  a large sample collected in 2003~\cite{pimumu} (see Figure \ref{mueff}).
A consistent result was obtained by considering $K^0 \rightarrow \pi^{\pm} \mu^{\mp} \nu$ 
decays collected in 2002 but with much lower statistics.
The resulting correction of $(+1.5\pm0.5)\%$ was applied in the BR calculation.

\begin{table}[htbp]
\begin{center}
\caption{Sources of systematic uncertainties.\label{systematics}}
\begin{tabular}{| l | r |}
\hline
Source & Uncertainty \\
\hline
 
 Background                        & $\pm6.4\%$ \\
 Normalization                     & $\pm3.0\%$ \\ 
 L2 trigger efficiency             & $\pm3.0\%$ \\
 Form factors                      & $\pm1.5\%$ \\
 Radiative corrections             & $\pm1.0\%$ \\
 Muon reconstruction efficiency    & $\pm0.5\%$ \\
\hline
 Total                             & $\pm7.9\%$ \\ 
\hline
\end{tabular}
\end{center}
\end{table}

\section{Conclusion}
Using data collected in 2002 with the NA48 detector at CERN, we 
obtain clear evidence of the decay 
$\Xi^{0}\rightarrow \Sigma^{+} \mu^{-} \overline{\nu}_{\mu}$, with
a precision on the branching ratio being significantly better than the existing
published value:
\begin{eqnarray}
\mathrm{BR}(\Xi^0 \rightarrow \Sigma^+ \mu^- \overline{\nu}_{\mu}) =(2.17 \pm 0.32_{\mathrm{stat}}\pm 0.17_{\mathrm{syst}})\times10^{-6}.
\end{eqnarray}
This result is in good agreement with the branching ratio measured by the 
NA48/1 collaboration for the electron channel, once the theoretical ratio of the corresponding 
decay amplitudes is taken into account~\cite{mendieta}.

\section*{Acknowledgments}
It is a pleasure to thank the technical staff of 
the participating laboratories,
universities and affiliated computing centres for their efforts in the 
construction of the NA48 apparatus, in the operation of the experiment, and in 
the processing of the data.

\newpage




\begin{thebibliography}{99}
%
%






\bibitem{cabibbo}
N.~Cabibbo, Physical Review Letters 10 (1963) 531.

\bibitem{donoghue}
J.F.~Donoghue,B.R.~Holstein and S.W.~Klimt,
Phys. Rev. D 35: (1987) 934.

\bibitem{mendieta}
R.~Flores-Mendieta, A.~Garcia and G.~Sanchez-Colon,
Phys.Rev. D54 (1996) 6855.

\bibitem{schlumpf}
F.~Schlumpf, 
Phys. Rev. D 51 (1995) 2262.

\bibitem{krause}
A.~Krause, 
Helv. Phys. Acta 63 (1990) 3.

\bibitem{anderson}
J.~Anderson and M.A.~Luty, 
Phys. Rev. D 47 (1993) 4975.

\bibitem{electron} J.R.~Batley et al. (NA48/1 Collaboration), Physics Letters B 645 (2007) 36.

\bibitem{ktevmu}
E.~Abouzaid et al. (KTeV Collaboration), 
Physical Review Letters 95 (2005) 081801.

\bibitem{detector} V.~Fanti et al. (NA48 Collaboration), Nuclear. Instrum. Meth. A 574 (2007) 433.


\bibitem{geant3} GEANT Description and Simulation Tool, CERN Program Library Long Writeup, W5013 (1994) 1.

\bibitem{Uli}
A.~Kadeer, J.~G.~Korner and U.~Moosbrugger, Eur. Phys. J. C 59 (2009) 27.

\bibitem{Gold}
M.~L.~Goldberger, S.~B.~Treiman, Physical Review 111 (1958) 354.

\bibitem{Comm}
E.~D.~Commins, P.~H.~Bucksbaum, ``Weak interactions of leptons and quarks'',
Cambridge University Press (2001).

\bibitem{ktevff}
A.~Alavi-Harati et al. (KTeV Collaboration), 
Physical Review Letters 87 (2001) 132001.

\bibitem{PDG} J.~Beringer et al. (Particle Data Group), Phys. Rev. D 86, (2012) 010001.

\bibitem{pimumu} J.~R.~Batley et al. (NA48/2 Collaboration), Phys. Lett. B 697, (2011) 107.








\end{thebibliography}
\end{document}